# Identification of Voice Utterance with Aging Factor Using the Method of MFCC Multichannel

Roy Rudolf Huizen[(1),(2)], Jazi Eko Istiyanto[(1)], Agfianto Eko Putra[(1)]

[(1)]Department of Computer Science and Electronics, FMIPA,
Universitas Gadjah Mada, (UGM) Yogyakarta, Indonesia.
[(2)]Computer System, STMIK STIKOM-Bali, Denpasar Bali, Indonesia.

*Abstract—* This research was conducted to develop a method to identify voice utterance. For voice utterance that encounters change caused by aging factor, with the interval of 10 to 25 years. The change of voice utterance influenced by aging factor might be extracted by MFCC (Mel Frequency Cepstrum Coefficient). However, the level of the compatibility of the feature may be dropped down to 55%. While the ones which do not encounter it may reach 95%. To improve the compatibility of the changing voice feature influenced by aging factor, then the method of the more specific feature extraction is developed: which is by separating the voice into several channels, suggested as MFCC multichannel, consisting of multichannel 5 filterbank (M5FB), multichannel 2 filterbank (M2FB) and multichannel 1 filterbank (M1FB). The result of the test shows that for model M5FB and M2FB have the highest score in the level of compatibility with 85% and 82% with 25 years interval. While model M5FB gets the highest score of 86% for 10 years time interval.

*Keywords;* Multichannel, MFCC, filterbank, aging factor

## I. INTRODUCTION

A Conversation is a form of communication, set by words into sentences. A conversation can be recorded, documented record is used to identify a plot of an event [1]. Words within the conversation have features, for each individual, those features are different from one to another [2],[3]. These features are obtained through extraction process; the method used are MFCC (Mel-Frequency Cepstral Coefficients) [4]. The features of the extraction result are the features of frequency [5]. For extraction mode of MFCC, it has Mel scale which constitutes a scale that has linear value for frequency under 1 KHz, and the exponential above 1 KHz.

The characteristics of frequency produced from the extraction consist of the fundamental frequency and formant frequency [6]. This characteristic can be used for identification by comparing the characteristics of voice utterance [7]. The compared characteristics are originated from the words in voice utterance. Similarity means that both are originated from the same individual and in contrary for the different characteristics [8].

The result of identification using extraction method of MFCC has the high level of compatibility and it can reach up to more than 95% [9]. The high compatibility can be reached for voice utterance that does not encounter change on its characteristics [8]. Some of these changes are caused by aging factor [10], [11]. The emergence of aging factor is because the occurrence of the time interval between voice utterance that will be identified and the voice utterance as the comparison [12], [13]. Time interval up to 25 years old causes the characteristics to change which is caused by the difference of age. The aging factor according to [6] and [13] can be observed to encounter change during a period of time of change between the age of 18 and 60 years old as shown in Figure 1. However, the existence of aging factor does not cause all characteristics to change yet only some components of characteristics. For components of characteristics that encounter change according to [11] is at the fundamental frequency and some formant frequency.

This research is to identify voice utterance with aging factor of the time interval of 10 and 25 years old. Identification using suggested extraction method, by developing a method of MFCC. This method of extraction is begun with separating voice into some range of frequency and the process of extraction is conducted in each channel. The process aims to obtain the more specific characteristics.

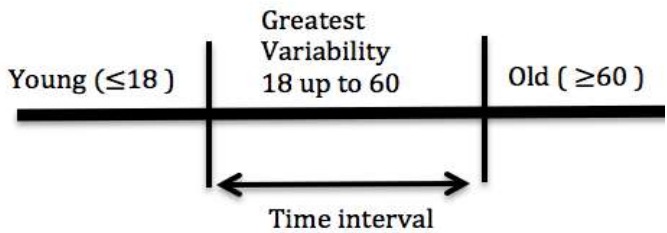

Figure 1 Time Interval Greatest Variability

## II. RELATED WORK

Extraction according to [14] is a process to emerge the characteristics of words in a recorded voice utterance. The characteristics consist of fundamental frequency (F0), and formant frequency. As for formant frequency, it is divided into formant frequency 1 (F1), formant frequency 2 (F2), formant frequency 3 (F3), and formant frequency 4 (F4) [11], [15]. According to [16], formant frequency has dynamic characteristics yet it can be used for identification process. Besides, based on research [15] the process of identification for voice that encounters language change can still be able to be analyzed using formant 1 up to formant 4 (F1 up to F4). The characteristics of voice with certain languages or accents [17] can be used to improve the performance of identification process of voice utterance recognition.

The characteristics of the voice utterance may encounter change [18], some of them are caused by the influence of noise interference [19]. Besides, it can also be caused by a conversation that occurs in a high tension, under alcohol consumption [18], [20], as well as the influence of aging factor [13], [10]. For the characteristics change caused by aging factor, occurs because of the existence of change in organ of voice in a particular time interval. The time interval according to [13] occurs between 18 and 60 years old. The research about the influence of aging on verification process such done by [13], states that the result of verification is highly influenced by the time interval between the taking of sample record and at the time of process of verification. In the research done by [11] states that the time interval influences the fundamental frequency value (F0) and the first formant value (F1) however there is no systematic effect on the second formant value (F2) and third formant (F3).

The analysis of the characteristics for voice utterance as done by [6] by dividing some range of frequency, the result of the research aims to recognize the age and gender obtain high accuracy up to 92.86%.

## III. RESEARCH METHODOLOGY

This research aims to identify the voice utterance influenced by aging factor, by using extraction method of MFCC multichannel. The stage of this method as shown in Figure 2 is begun with the process of pre-emphasis as shown on the (1). The function of pre-emphasis, so that the spectrum will be smoother, more even or flat. The output of pre-emphasis is influenced by value for model of suggested value used is 0.97 [12].

$$Sout(n) = Sin(n) - âSin(n-1) \quad (1)$$

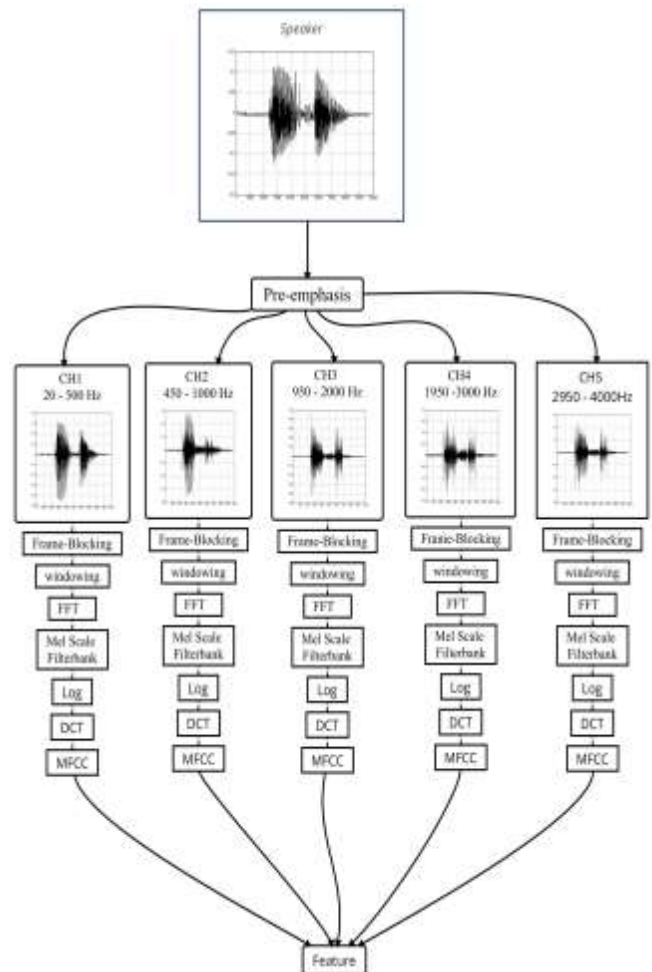

Figure 2 Block diagram of MFCC multichannel method

The output of pre-emphasis is divided into some channels of frequency using a filter FIR (finite impulse response). The division into some ranges of frequency to obtain more specific characteristics. Based on the characteristics of the fundamental frequency and formant, therefore, channel division is divided into 5 (Ch1, Ch2, Ch3, Ch4, and Ch5). For Ch 1 using filter FIR has bandwidth of 20 Hz up to 500 Hz, Ch2 with the range of 450 Hz up to

1000 Hz, Ch3 950 up to 2000 Hz, Ch4 1950 Hz up to 3000 Hz and Ch5 with the range of 2950 Hz up to 4000 Hz. The Bandwidth of the filter commonly uses (2). For Ch1 to form filter LPF, Ch2 filter BPF1, Ch3 filter BPF2, Ch3 filter BPF3 and Ch5 filter BPF4. Range bandwidth from filter FIR is determined by filter response (h[k]) and b is the constantan of filter [21].

$$y[n] = \sum_{k=0}^{M} b_k S_{out[n-k]} = \sum_{k=0}^{M} h[k] S_{out[n-k]} \quad (2)$$

For output from each channel is ych1(n), ych2(n), ych3(n), ych4(n), and ych5(n), are divided into the time domain to become several frames using frame-blocking. The width of frame-blocking used is 20 ms [13] with the overlapping is 5 ms. To divide the voice in each channel, (3) is used.

f(1;n) = s' (n+M(ℓ-1))

n= 0,….N-1,  ℓ = 1:L    (3)

Each voice frame in each channel is windowing hamming [22] and is counted the frequency using FFT. The frequency value obtained is passed through the Mel scale using filterbank. The Mel scale has similar characteristics as human hearing which linear under 1000 Hz and nonlinear above 1000Hz. Therefore, in order to make the process of extraction can be compatible with the characteristics, the process of adjustment is needed using Mel Scale. To convert it from the output of FFT becomes Mel scale using equity (4). While to convert it back to Mel scale to the linear frequency, Equity (6) is used. The equity of Mel scale in each channel is shown in equity (5). The width of Mel scale is ruled based on the range from the frequency of filter FIR.

$$f_{Mel} = 2595 * \log_{10}\left[1 + \frac{f_{linear}}{700}\right]$$

or

$$f_{Mel} = 1127 \left[1 + \frac{f_{linear}}{700}\right] \quad (4)$$

Each channel uses the following equity;

$$f_{Mel\_chn} = 2595 * \log_{10}\left[1 + \frac{fchn}{700}\right]$$

or

$$f_{Melchn} = 1127 \left[1 + \frac{f_{chn}}{700}\right] \quad (5)$$

CHn consists of Ch$_1$, Ch$_2$, Ch$_3$, Ch$_4$ and Ch$_5$. To regain back the linear frequency of each channel, the following (6) is use;

$$f_{linear}^{-1} = 700 \left[10^{\frac{fmel}{2595}} - 1\right]$$

or

$$f_{linear}^{-1} = 700 \left[e^{\frac{fmel}{1127}} - 1\right] \quad (6)$$

Results of the filterbank, inverse (log DCT) to get the value MFCC, in the each channel. Output for each channel has characteristics that need to analyze for identification. The following is the result of extraction on each channel in Figure 3.

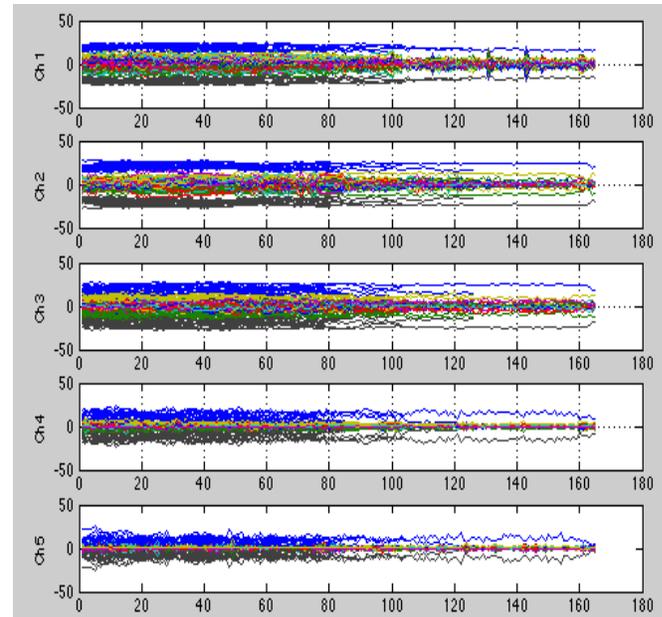

Figure 3 The Result of Extraction of Each Channel

Model extraction of MFCC-MultiChannel for filterbank part is used 4 as shown in Figure 4. for Figure 4(a), filterbank in each channel has the different range. For Figure 4(b), it is divided into two ranges of frequency: under 1 KHz and above 1KHz. For the one under 1 KHz it is divided into 2 channels while the above 1 KHz is divided into 3 channels. In channel 1 and 2 the number of filterbanks used each is 18 and 15. While for channel 3, 4, and 5 each uses as many filterbanks as 12, 10 and 8. For Figure 4(c) the number of filterbanks in each channel is 18, 15, 12, 10 and 8 with the range of each channel is as much as 20 to 4 KHz. For Figure 4 (d) constitutes a single extraction model channel, the number of filterbanks used is as much as 22. The result of extraction for model M5FB, M2FB and M1FB and SCFB each is analyzed to obtain maximum value, minimum value, standard deviation (sd) and mean.

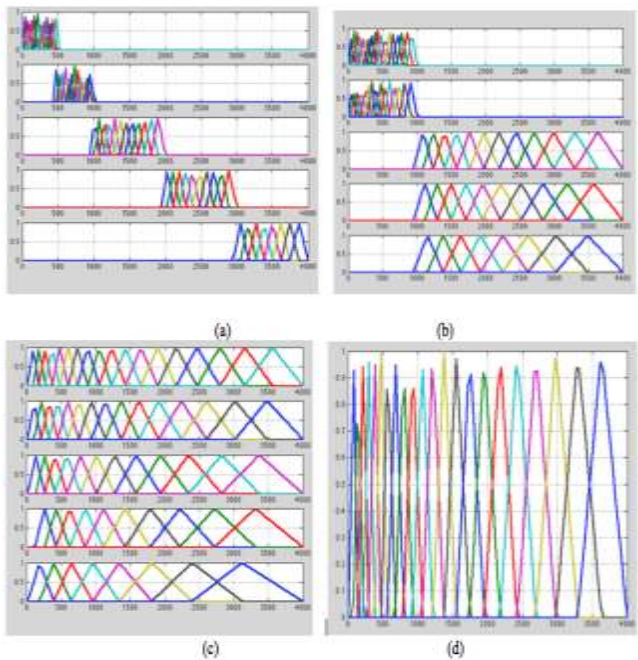

Figure 4 Filterbank Variants (a) multichannel 5 FilterBank (M5FB), (b) multichannel 2 FilterBank (M2FB), (c) multichannel 1 FilterBank (M1FB) dan singlechannel FilterBank (SCFB)

## IV. RESULTS

The test using model M5FB, M2FB, and M1FB and SCFB for data identification on Figure 5. The test for time interval 10 and 25 years old, and also 1 year old. 20 words are taken from each recording to be extracted and obtain the characteristics [20]. The type of word used is text independent.

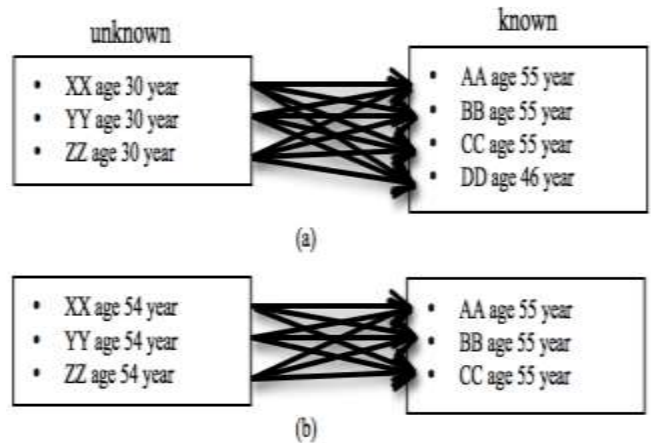

Figure 5 Data test

Each characteristic of the words that has been extracted is compared, the pattern of the characteristic observed is the max value, min value, standard deviation and mean. The result of extraction from the recorded voice utterance XX compared to the recorded voice utterance BB and CC. Between XX and BB or CC have the time interval of 25 years, the result of extraction is shown in Figure 6.

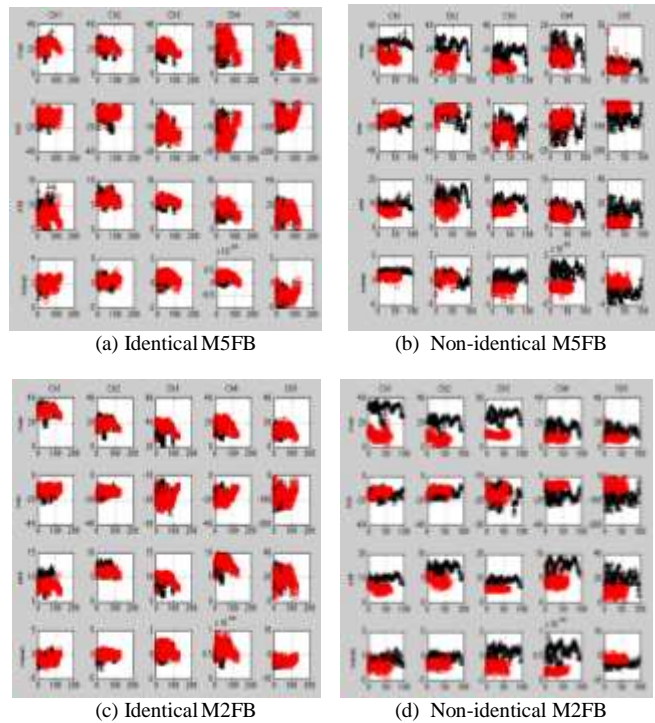

(a) Identical M5FB        (b) Non-identical M5FB

(c) Identical M2FB        (d) Non-identical M2FB

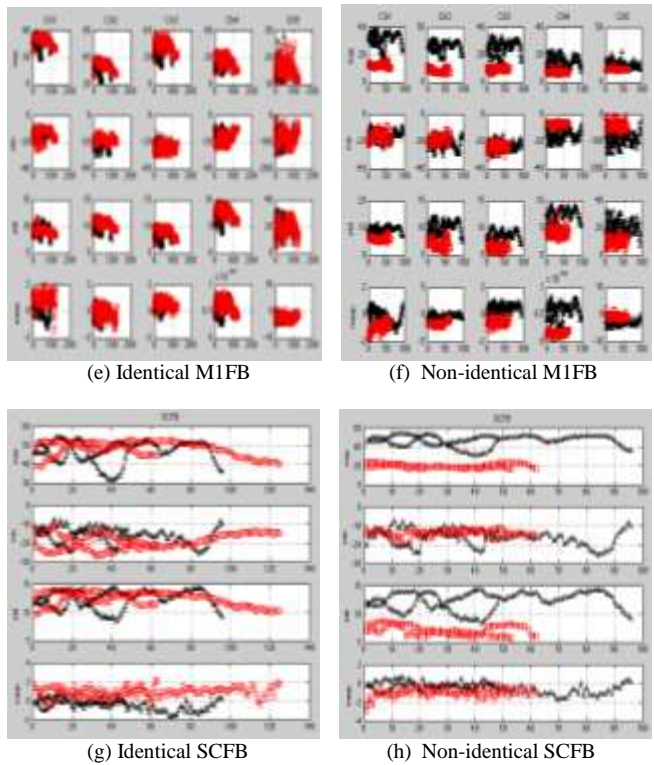

(e) Identical M1FB    (f) Non-identical M1FB

(g) Identical SCFB    (h) Non-identical SCFB

Figure 6 Comparing the result of extraction with interval of 25 years using M5FB, M2FB, M1FB and SCFB

For each similar pattern of characteristics are considered that both individuals are identical while for the different pattern of characteristic is stated as nonidentical. The result of test using various variant of filterbank is shown in Table 1.

Table 1 Test Result

| Interval | M5FB | M2FB | M1FB | SCFB |
|---|---|---|---|---|
| 25 year | 85% | 82% | 72% | 55% |
| 10 year | 86% | 82% | 80% | 70% |
| 1 year | 95% | 95% | 95% | 95% |

In Table 1 for interval 25 years with the variant of M5FB and M2FB, the highest score of 85% and 82 % level of compatibility is obtained. For interval 10 years, variant M5FB obtained the highest score of 86% as well as for interval 1 year with 95% compatibility value. While in single filterbank the highest accuracy is in interval 1 year, and the lowest is in interval 25 years with 55% (SCFB).

## V. CONCLUSION

Model extraction of suggested MFCC-MultiChannel for M5FB in its compatibility the result is 85%. While for SCFB the level of compatibility for interval 25 years is 55%. The different result of this test shows that the division of voice utterance becomes several ranges of frequency is proven to be able to be used to gain better characteristics compared to without separating then into several channels.

## VI. ACKNOWLEDGMENT

This research is supported by Doctorate Program in Computer Science, Department of Computer Science and Electronics, Faculty of Mathematics and Natural Sciences, Universitas Gadjah Mada, and Computer System, STMIK STIKOM-Bali We would like to acknowledge for their support in this research.